\documentclass[showpacs,preprint,aip]{revtex4}
\usepackage{graphicx}
\usepackage{dcolumn}
\usepackage{bm}
\usepackage{color}

\begin{document}
\title{Nearly-zero transmission through periodically modulated ultrathin metal films}
\author{Sanshui Xiao$^1$}
\email{saxi@fotonik.dtu.dk}
\author{Jingjing Zhang$^1$}
\author{Liang Peng$^1$}
\author{Claus Jeppesen$^2$}
\author{Radu Malureanu$^1$}
\author{Anders Kristensen$^2$}
\author{N. Asger Mortensen$^1$}
\affiliation{$^1$ DTU Fotonik - Department of Photonics Engineering, \\Technical
University of Denmark, DK-2800 Kongens Lyngby, Denmark.\\
$^2$ DTU Nanotech - Department of Micro and Nanotechnology, Technical University of Denmark,
DK-2800 Kongens Lyngby, Denmark.}

\date{\today}

\begin{abstract}
Transmission of light through an optically ultrathin metal film with a thickness comparable to its skin depth is significant. We demonstrate experimentally nearly-zero transmission of light through a film periodically modulated by a one-dimensional array of subwavelength slits. The suppressed optical transmission is due to the excitation of surface plasmon polaritons and the zero-transmission phenomenon is strongly dependent on the polarization of the incident wave.
\end{abstract}
\maketitle



By the presence of periodic subwavelength holes, an otherwise optically opaque metal film (i.e. thickness larger than 100~nm) can pass extraordinary amounts of light at certain frequencies, so-called extraordinary optical transmission, which was firstly reported by Ebbesen \emph{et al.}\cite{Ebbesen1998P1}. The mechanisms behind it have been widely addressed and it has been shown that the enhanced transmission is mostly attributed to the surface plasmon polaritons (SPPs)~\cite{Ghaemi1998P1,Porto1999P1,Martin2001P1,Xiao2007P1,Garcia2005P2,Xiao2010P1}.
On the contrary, for an ultrathin metal film with the thickness comparable to the optical penetration depth, it is well known that the transmission of light through it becomes prominent~\cite{JDJackson}.
Thus, to much surprise it was recently predicted by means of analytical and numerical calculations~\cite{Spevak2009P1} that the transmission through an ultrathin film was suppressed when periodically modulated. Both a quantitative and qualitative explanation of the effect was offered in Ref.~\cite{Spevak2009P1} and the case of two-dimensional periodicity was emphasized as well. Subsequently, the suppressed transmission through an ultrathin film with a two-dimensional periodic array of subwavelength holes was experimentally demonstrated~\cite{Braun2009P1}.
In this Letter, we demonstrate experimentally nearly-null transmission of light through semi-transparent metal films modulated by one-dimensional periodic arrays of subwavelength slits. The suppressed optical transmission is attributed to the excitation of surface plasmon resonances. Influence of the structure size to the resonance is systemically investigated.

\begin{figure}[h]
\includegraphics[width=6.0in]{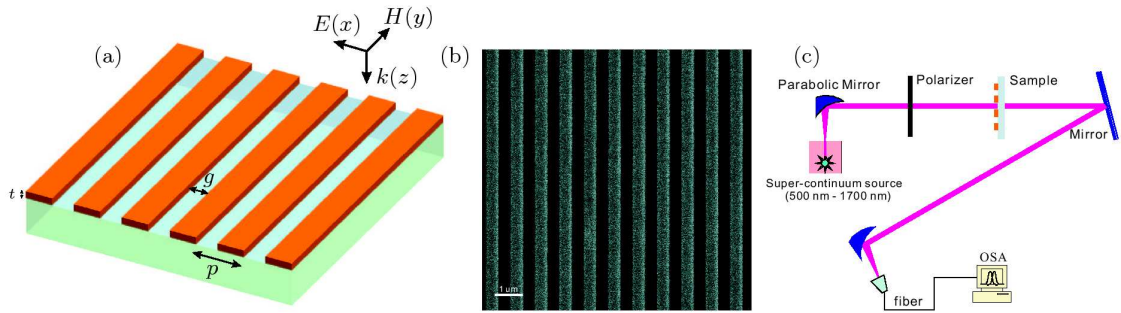}
\caption{\label{structure} (Color online) (a) A schematic illustration of a one-dimensional periodic array of subwavelength slits perforated in
an ultrathin metal film on a quartz wafer, where the thickness of the film is $t$ = 15~nm, the width of the slits is $g$ = 400~nm, and the period of the slit array is $p$ = 900~nm. Axis shows the propagation direction and polarization of the incident plane wave. (b) Scanning electron microscope image of its corresponding structure. (c) Free-space measurement setup.}
\end{figure}

\begin{figure}[h]
\includegraphics[width=3.0in]{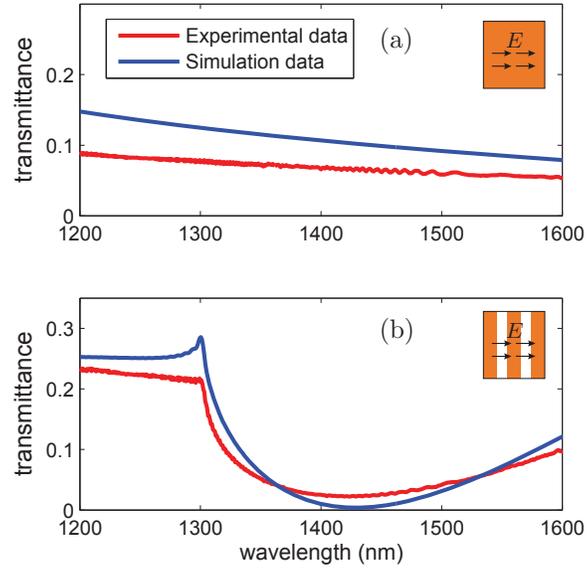}
\caption{\label{spectra} (Color online) Transmittance spectra for the closed (a) and perforated (b) ultrathin gold film at normal incidence. The red lines correspond to the experimental results and the blue lines are the results from numerical analysis. Insets show the polarization of the electric field for the incident waves.}
\end{figure}

The perforated gold film, i.e. the one-dimensional grating structure, was fabricated on a fused silica substrate by electron beam lithography (EBL). Firstly,  a 100~nm ZEP520A resist was spincoated onto the silica substrate. Then 15~nm aluminium was deposited on top of the ZEP520A layer to prevent charging during EBL. Following exposure, the resist was developed in ZED-N50. Then 5~nm titanium and 15~nm gold was deposited before lift-off was performed concluding the fabrication process. A similar nanofabrication procedure is described in detail in Ref.~\cite{Claus2009P2}. A schematic illustration of the one-dimensional grating structure and the scanning electron microscopy image are shown in Fig.~\ref{structure}, where the thickness of the film is $t$ = 15~nm, the width of the slits is $g$ = 400~nm, and the period of the array is $p$ = 900~nm. The inset in Fig.~\ref{structure} (a) indicates the propagation direction and polarization of the incident plane wave. Here we consider that light is normally incident on the sample and the electric field of the incident wave is perpendicular to the slits. The sample was characterized by free space setup, illustrated in Fig.~\ref{structure} (c). Transmission through the modulated structure was compared with the transmission through a closed, i.e. non-modulated, 15~nm gold film.
The red lines in Fig.~\ref{spectra} (a) and Fig.~\ref{spectra} (b) represent the experimental results of the transmittance spectra for the closed and modulated ultrathin metal films at normal incidence, respectively. At the wavelength of our interest as shown in Fig.~\ref{spectra}, one can observe that around 10\% of the intensity is transmitted through the closed metal film with the thickness of 15~nm, as shown in Fig.~\ref{spectra} (a). When periodically modulated by an array of one-dimensional slits, transmission is dramatically suppressed.  Especially for the wavelength around 1428.6~nm, the transmittance decreases significantly from 6.7\% to 2.1\%. Intuitively one expects that the metal film with slits, compared with the closed film, would transmit more light, since less material is blocking the light. However, we emphasize that the measurement demonstrates the opposite: less light is transmitted through the perforated film! This nontrivial effect is also verified by numerical simulations, shown by the blue lines in Fig.~\ref{spectra}. Note that we used the experimental data from Ref.~\cite{Johnson1972P1} for the optical properties of gold and the refractive index of the quartz is 1.445.

\begin{figure}[h]
\includegraphics[width=5.0in]{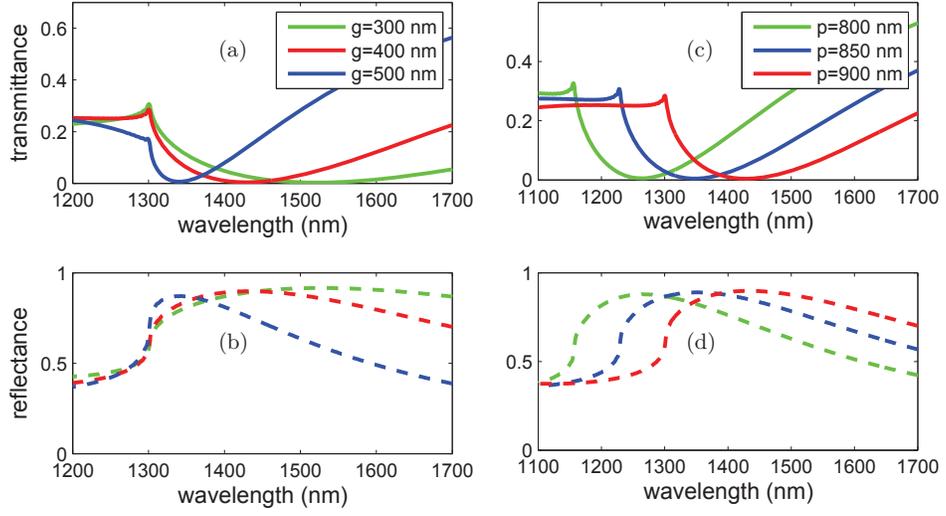}
\caption{\label{parameterstuning} (Color online) Transmittance (a) and reflectance (b) spectra when varying the gap $g$ for $p$ = 900~nm.
Transmittance (c) and reflectance (d) spectra as a function of the periodicity $p$ of the structured film for $g$ = 400~nm.}
\end{figure}

The wavelength 1428.6~nm associated with the suppressed transmission in Fig.~\ref{spectra} exceeds corresponding Rayleigh values ($\lambda_{\rm Rayleigh} = p \sqrt{\varepsilon}$). In Fig.~\ref{spectra} we can observe the Rayleigh anomaly at the wavelength 1300~nm.
We emphasize that the suppressed transmission at 1428.6~nm is attributed to the excitation of the surface plasmon polariton at the quartz/gold interface. The surface plasmon resonance ($\lambda_{\rm spp}$) at the interface (without slits) of the quartz/gold semispaces is around 1322~nm.
Due to ultrathin-thickness of the film and the presence of the slits, the resonant wavelength is redshifted to 1428.6~nm. In fact, we also observe the resonance around 970~nm (not illustrated in Fig.~\ref{spectra}), which we attribute to the excitation of the surface plasmon polariton at the air/gold interface.
Figure~\ref{parameterstuning} shows the transmittance and reflectance when varying the size of the slits $g$ and the period of the structure $p$.
The resonant wavelength associated with the suppressed optical transmission is blueshifted with increasing $g$. When enlarging $p$, the resonance is redshifted. From Fig.~\ref{parameterstuning}(b), we can see that most of the energy, at the resonant wavelength, is reflected instead of being absorbed.
For the structured film at the resonance $\lambda$ = 1428.6~nm, the transmittance ($T$), reflectance ($R$), and absorption ($A$) are 0.39\%, 89.8\%, and 9.81\%, respectively. Note that $T$ = 10.21\%, $R$ = 85.16\%, and $A$ = 4.63\% for the case of the closed film. One can conclude that the excitation of the SPPs results in the suppression of the transmittance and increase of the reflectance as well as the absorption.
We can also observe from Figs.~\ref{parameterstuning} (a) and ~\ref{parameterstuning} (c) that the Rayleigh anomaly is independent on $g$, while strongly dependent on $p$. The wavelengths associated with the Rayleigh anomaly are 1156~nm, 1228~nm, and 1300~nm for $p$ = 800~nm, $p$ = 850~nm, and $p$ = 900~nm, respectively. The results agree well with what we expected since the values with the Rayleigh anomaly are equal to the product of the grating period and the refractive index of the substrate.

\begin{figure}[h]
\includegraphics[width=3.0in]{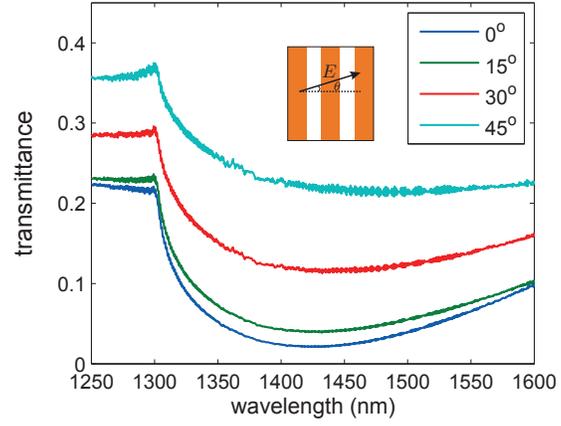}
\caption{\label{spectra-angle} (Color online) Measured transmittance spectra of the perforated film when varying the angle of polarization $\theta$.}
\end{figure}

Finally we demonstrate experimentally the influence of the polarization angle on the zero-transmission effect. Notice that the results mentioned above are all for the case when the electric field of the incident wave is perpendicular to the slits ($\theta=0$). Figure~\ref{spectra-angle} illustrates the transmittance when varying the angle of polarization $\theta$. The suppressed transmission phenomenon clearly weakens when increasing $\theta$.
As discussed above, the suppressed transmission is attributed to the excitation of the SPP, which is a transverse magnetic wave $\mathbf{H}_{\rm spp} = A {\mathbf{\hat{y}}}$ with the wave vector $\mathbf{k}_{\rm spp}=2\pi/p{\mathbf{\hat{x}}}$.
Therefore, the excitation of the SPP is mainly governed by the value of the y-component of the incident magnetic field. The larger the field strength is, the stronger the SPP excitation.
The transmittance contributed from the SPP resonance is proportional to $cos^2(\theta)$. Likewise, the contribution to the transmittance from the orthogonal component of the incident magnetic wave is proportional to $sin^2(\theta)$. Consequently, the energy flux of the transmitted
wave includes both contributions and is proportional to
$|T_1|^2sin^2(\theta)+ |T_2-T_r|^2cos^2(\theta)$, where
$T_1$ ($T_2$) is the non-resonant transmittance coefficient
relating to the incident magnetic field component perpendicular
(tangential) to the slits and $T_r$ denotes the resonance coefficient relating to the SPP
excitation. 

In summary, we have experimentally demonstrated nearly-zero transmission of light through the ultrathin metal film (with the thickness in the order of its skin depth), which is periodically modulated by a one-dimensional periodic array of subwavelength slits.
Supported by numerical simulations, we have shown that the suppressed optical transmission is associated with the excitation of standing surface plasmon-polariton waves. Experimental results show that this zero-transmission effect is strongly dependent on the polarization of the incident wave. The structured ultrathin film may be applied to create mirrors for laser devices, small polarizing filters and other components for photonic devices.

This work is partly supported by the Danish Research Council for Technology and Production Sciences (grant no: 274-07-0379) and
the Catalysis for Sustainable Energy initiative Center, funded by the Danish Ministry of Science, Technology and Innovation. We acknowledge stimulating discussions with an anonymous Referee.


\newpage
\section{Figure captions}
\textbf{Figure 1}: (Color online) (a) A schematic illustration of a one-dimensional periodic array of subwavelength slits perforated in
an ultrathin metal film on a quartz wafer, where the thickness of the film is $t$ = 15~nm, the width of the slits is $g$ = 400~nm, and the period of the slit array is $p$ = 900~nm. Axis shows the propagation direction and polarization of the incident plane wave. (b) Scanning electron microscope image of its corresponding structure. (c) Free-space measurement setup.

\textbf{Figure 2}: (Color online) Transmittance spectra for the closed (a) and perforated (b) ultrathin gold film at normal incidence. The red lines correspond to the experimental results and the blue lines are the results from numerical analysis. Insets show the polarization of the electric field for the incident waves.

\textbf{Figure 3}: (Color online) Transmittance (a) and reflectance (b) spectra when varying the gap $g$ for $p$ = 900~nm.
Transmittance (c) and reflectance (d) spectra as a function of the periodicity $p$ of the structured film for $g$ = 400~nm.

\textbf{Figure 4}: (Color online) Measured transmittance spectra of the perforated film when varying angles of polarization $\theta$.

\newpage
\clearpage

\newpage
\clearpage

\newpage
\clearpage

\newpage
\clearpage

\end{document}